\documentclass[twocolumn,runningheads]{svjour2}
\smartqed  
\usepackage{graphicx}
\def\psrb{PSR~B1259--63 } 
\def\gr{$\gamma$-ray}
\def\xmm{\textit {XMM-Newton}}
\def\sax{\textit{Beppo}SAX}
\def\asca{\textit{ASCA}}
\journalname{Astrophysics and Space Science}
\begin{document}

\title{Rradio-to-TeV \gr\ emission from  \psrb}


\author{Andrii Neronov \and Maria Chernyakova 
}

\institute{ISDC, Ch. d'Ecogia 16, 1290 Versoix, Switzerland\\
              \email{andrii.neronov@obs.unige.ch} \\
}

\date{Received: date / Accepted: date}

\maketitle

\begin{abstract}
We discuss the implications of the recent X-ray and TeV \gr\ observations 
of the \psrb\ system (a young rotation powered pulsar orbiting a Be star)
for the theoretical models of interaction of pulsar and stellar winds.
We show that previously considered models have problems to account for the observed behaviour of the system. 
We develop a model in which the broad band emission from the binary system is produced in
result of collisions of  GeV-TeV energy protons accelerated by the pulsar wind
and interacting with the stellar disk. In this model the  high energy \gr s are
produced in the decays of secondary neutral pions,  while radio and X-ray  emission  are
synchrotron and inverse Compton emission produced by low-energy ($\leq 100$
MeV)  electrons from the decays  of secondary  charged $\pi^\pm$ mesons. 
This model can
explain  not only the observed energy spectra,  but also the correlations
between TeV, X-ray  and radio  emission components.  
\end{abstract}

\section{Introduction}
\label{intro}

\psrb{} is a $\sim$48 ms radio pulsar in a highly eccentric (e$\sim$0.87), 3.4
year orbit with a Be star SS 2883 \cite{johnston92}. The pulsar crosses the
Be star disc twice per orbit, just prior to and just after periastron. Unpulsed
radio, X-ray and \gr\ emission observed from the binary system are produced due
to the collision of pulsar wind with the wind of Be star. Observations of the 
temporal and spectral evolution of the non-thermal emission from the system 
provide a unique opportunity to probe the physics of the pulsar winds (PW) 
which is, in spite of the wealth of observational phenomena, and a 40-year old
observation history, remains a matter of debate. 

The interaction of the PW with the
wind from the companion star, is responsible for the formation of a 
"compactified" pulsar wind nebula (PWN) with the the size about the binary
separation distance (typically, on AU-scale).   Compact size, large matter
density and the presence of a strong source (companion star) which 
illuminates  the nebula make the physical properties of the compact PWN
significantly different from the ones of their larger scale cousins.

We present the results of the last observation campaign of
the \psrb\ system during the 
2004 pulsar periastron passage and their applications for the theoretical modelling of the source. We show
that most of the observed properties of the system in radio-to-TeV band can be
naturally explained within a model of proton-loaded pulsar wind.

\section{Multi-wavelength observations of the system during 2004 periastron
passage.}

\begin{figure}
\centering
\includegraphics[width=\columnwidth,angle=0]{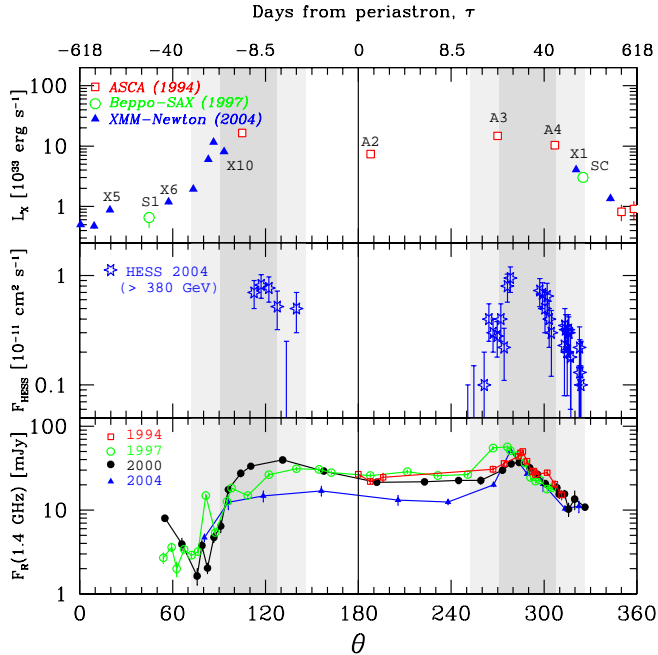}
\caption{Comparison between the X-ray (top), TeV (middle) and radio
  (bottom) lightcurves.  \xmm\ observations are marked with triangles,
  \sax\ ones with circles, and \asca\ ones with squares. Data for four
  different periastron passages. are shown with different colors: red
  (1994), green (1997), black (2000) and blue (2004). Bottom X axis
  shows the orbital phase, $\theta$, top X axis shows days from
  periastron, $\tau$.}
\label{Xradio}
\end{figure}
The upper panel of Fig. \ref{Xradio} shows the X-ray lightcurve of the
system \cite{chernyakova06} together with the TeV \cite{aharonian05} and radio
\cite{johnston05} lightcurves. For comparison we show also the data from
archival X-ray \cite{kaspi95,hirayama99} and radio \cite{johnston99,connors02} observations.
 Rapid growth of the X-ray flux found in  \xmm\
observations of 2004 is correlated with the rapid growth of the 
unpulsed radio emission from the system. The growth of radio
and X-ray flux at these phases can be attributed to the pulsar
entering the Be star disk.  

Unfortunately, TeV observations start somewhat later and it is not
possible to see whether the TeV flux grows during the pre-periastron
disk crossing. However, simple geometrical argument tells that the
orbital phase $\theta$ at which the pulsar should enter the disk for
the second time should be shifted by $180^\circ$ relative to the first
entrance.  From Fig.  \ref{Xradio} one can infer that the first
pre-periastron entrance falls roughly between the phases
$70^\circ<\theta<110^\circ$. Thus, the pulsar has to enter the disk
again between the phases $250^\circ<\theta<290^\circ$. Surprisingly,
one can clearly see from the middle panel of Fig. \ref{Xradio} that
the TeV flux grows in this phase interval. 
\begin{figure}
\centering
\includegraphics[width=7cm,angle=0]{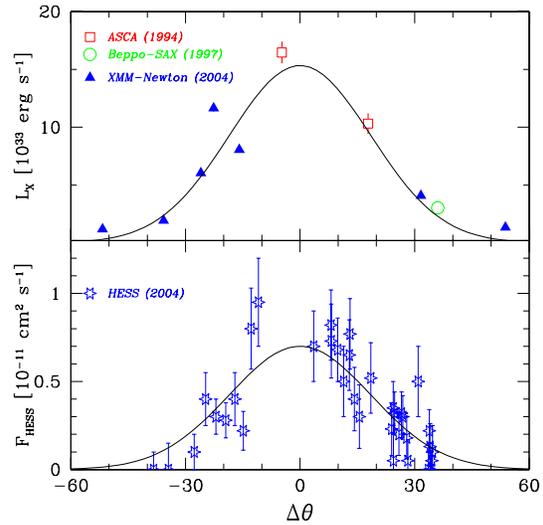}
\caption{The X-ray (top) and TeV (bottom) flux as a function of the relative
  phase $\Delta\theta=\theta-\theta_0$ (see text for the
  definition of $\theta_0$). The curves show a fit with a gaussian of the
  half-width $\Delta\theta_0=18.5^\circ$.}
\label{gauss}
\end{figure}
To test the conjecture that the TeV flux grows during the second entrance to the disk
we superimpose the pre-periastron X-ray and
TeV lightcurves over the post-periastron lightcurves by shifting the
phase of the post-periastron data points by $-180^\circ$. The result
is shown in Fig. \ref{gauss}. One can see that in such representation
the rise and decrease of both X-ray and TeV flux from the system can
be well fitted with a gaussian curve
$F(\theta)\sim\exp\left(-(\theta-\theta_0)^2/(2\Delta\theta_0^2)\right)$.
We find that the best fit is achieved with the parameter choice
$\theta_0\simeq 109.1^\circ$, $\Delta\theta_0\simeq
18.5^\circ$ (the coordinate $\Delta\theta$ along the X-axis of
Fig. \ref{gauss} is, in fact $\Delta\theta=\theta-\theta_0$).
The position of the Gaussian with the above parameters is 
shown schematically in Fig. \ref{Xradio} by a shaded area.
Denser and lighter  shadings in Fig. \ref{Xradio} correspond to the one and two 
widths of the
Gaussian. Depending on the physical mechanisms of the X-ray and TeV emission, the inferred width of 
the Gaussian $\Delta\theta_0$ gives either an estimate of the ``thickness'' of the Be star disk, 
or of the characteristic cooling time of the high-energy particles injected at the phase of the disk passage (see Section \ref{sec:last}).

\begin{figure}
\centering
  \includegraphics[width=0.8\columnwidth,angle=0]{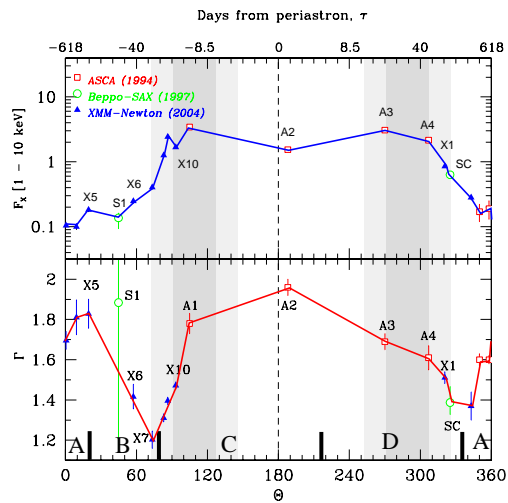}
\caption{Evolution of the X-ray photon index $\Gamma_{ph}$ over the orbital
phase $\theta$. Radio flux and spectral index evolution from the 1997 periastron passage
\cite{johnston99} are shown in black.}
\label{fig:Ph_index}       
\end{figure}

The graphical representation of the evolution of the X-ray photon index (the
spectrum is well fit by an absorbed powerlaw model)
along the orbit is given in Figure \ref{fig:Ph_index}.
The most remarkable feature of the spectral evolution of the system is
the hardening of the X-ray spectrum close to the moment when pulsar
enters the Be star disk at the phase $\theta\simeq
\theta_0-2\Delta\theta_0\simeq 70^\circ$. One can 
see that the decrease of the photon index $\Gamma$ is simultaneous
with the onset of the rapid growth of the X-ray flux. Similar hardening of the spectrum down to the photon index $\Gamma_{ph}\simeq 1$ (or, equivalently, down to the spectral index $\alpha\simeq 0$) at the moment of disk entrance is observed in the radio data shown in Fig. \ref{fig:Ph_index} in black \cite{johnston99}. 
 To the best of our
knowledge, neither the hardening of the X-ray spectrum, nor strong correlation between the radio and X-ray flux and spectral index variations was predicted in any of existing models
of X-ray emission from the system.  

\section{Implications for theoretical models.}

 Termination of the pulsar
wind in the stellar disk  leads to the acceleration  of ultrarelativistic
electrons and subsequent  X-ray and gamma-ray emission. Within a simple geometrical
picture \cite{tavani97}, the maximum of X-ray
(synchrotron) emission is expected  during the phases of the disk passage,
while the maximum of TeV \gr (IC) emission is expected at the moment of the
periastron passage. However, surprisingly, the TeV \gr\ lightcurve (middle panel
of Fig.
\ref{Xradio})  has a local minimum in the periastron.  The observed X-ray -- TeV and radio -- X-ray
correlations  do not agree  with the early theoretical predictions
\cite{tavani97,kirk99}. To find the range of possible
theoretical models which can explain the data it is useful first to make basic
qualitative estimates of different time scales present in the system.

\subsection{Characteristic cooling times.}
\paragraph{Electrons.}
One of the main differences between the synchrotron and IC mechanisms
of X-ray emission is the difference in the cooling time scales. 
 The cooling time of the TeV electrons
which can produce synchrotron emission at the energies 
$\epsilon_S\sim 1-10$~keV in the magnetic field $B$ is 
\begin{equation} 
\label{eq:ts}
t_{S}\simeq 6\times 10^2
\left[B/0.1\mbox{G}\right]^{-3/2}
\left[\epsilon_S/10\mbox{ keV}\right]^{-1/2}
\mbox{ s.} 
\end{equation}
The spectrum of optically thin synchrotron emission from the cooled electron
population has the photon index $\Gamma_{ph}\ge 1.5$. Any hardening of the X-ray
spectrum down to the values $\Gamma_{ph}<1.5$ (e.g. due to the increased
injection of electrons at higher energies) would be "washed out" by the
synchrotron cooling at the at the $10^2-10^3$~s time
scale. To the contrary, the typical IC cooling time in X-rays is
\begin{equation}
\label{eq:tic}
 t_{IC\rm\ (T)}\simeq 6\times 10^5\left[R/10^{13}\mbox{cm}\right]^{2}
\left[\epsilon_{IC}/10\mbox{ keV}\right]^{-1/2}\mbox{ s.}
\end{equation}
(we have assumed that the seed photons for the IC scattering come
from the companion star of luminosity $L_*\simeq 10^{38}$~erg/s and temperature
$T\simeq 2\times 10^4$~K; the subscript ``(T)'' indicates that the estimate applies for the Thompson regime). Estimating the size of emission region to be about the binary
separation distance, $R\sim 10^{13}$~cm, one can  find that
electrons emitting IC radiation at 1~keV cool at the day time scales. 
Observation of the gradual evolution of the X-ray photon index down to
$\Gamma_{ph}\simeq 1.2$ and then back to $\Gamma_{ph}\ge 1.5$ on the time scale of
several days during the first entrance to the disk (see Fig.
\ref{fig:Ph_index}) 
is consistent with the IC, rather than synchrotron model of X-ray emission. 

In principle, it is possible that X-ray and TeV $\gamma$-ray emission from the
system are, respectively, low- and high- energy tails of the IC
spectrum. Substituting naively the energy of TeV photons $\epsilon_{IC}\sim
1$~TeV into Eq. (\ref{eq:tic}) one finds that the time scale of the spectral
variability at TeV energies should be very short. However, at TeV energies the
IC scattering proceeds in the Klein-Nishina regime and the cooling time in this
regime grows with energy,
\begin{equation}
\label{eq:tickn}
t_{IC\rm\ (KN)}\simeq 8.5\times 10^3\left[\epsilon_{IC}/1\mbox{ TeV}\right]^{0.7}\left[R/10^{13}
\mbox{ cm}\right]^2\mbox{ s}
\end{equation}
The minimum of the IC cooling time, $\sim 10^3$~s, is reached at roughly at the energy of
transition between Thompson and Klein-Nishina regimes, $E_e\sim\epsilon_{IC}\sim
 10-100$~GeV.

Pulsar wind electrons are able to escape from the region of the dense photon
background along the contact surface of pulsar and stellar wind. If the two
winds do not mix,  the pulsar wind flows along the
contact surface with the speed $v_{\rm PW}\sim
10^{10}$~cm/s and  escapes beyond the binary separation distance over the time
scale  
\begin{equation}
\label{eq:esc}
t_{esc}\sim R/v_{\rm PW}\simeq
10^3\left[R/10^{13}\mbox{ cm}\right]\mbox{ s}
\end{equation}
This time scale is essentially shorter
than the IC cooling time both in X-ray and in the TeV energy bands. 
Electrons escaping from the innermost region of pulsar wind/stellar wind 
interaction fill the larger extended region (a "compactified" PWN
of the size $R_{\rm PWN}$ of about several binary separation distances) and can loose their energy via IC
emission at longer time scales in the less dense photon background produced by
the Be star. The escape time from the compact PWN can be naively estimated
assuming diffusion in the weak PWN magnetic field. E.g. taking the diffusion
coefficient $D$ equal to the Bohm diffusion coefficient at $E_e\sim 1$~TeV 
and depending on the energy as $D\sim E^{-\alpha}$ ($\alpha=1$ for the case of
Bohm diffusion) one finds
\begin{equation}
\label{eq:PWN}
t_{\rm PWN}\simeq 10^4\left[B/0.1\mbox{G}\right]\left[E_e/1\mbox{ TeV}
\right]^{-\alpha}\left[R_{\rm PWN}/10^{13}\mbox{ cm}\right]^2\mbox{ s}
\end{equation} 


During the periods of the pulsar passage through the dense equatorial
disk of Be star (typical density of the slow equatorial stellar wind at the
location of the pulsar is $n_{\rm disk}\sim 10^{10}-10^{11}$~cm$^{-3}$),
bremsstrahlung and ionisation energy losses can compete with the IC loss.
Indeed, the energy independent bremsstrahlung loss time,
\begin{equation}
\label{eq:brems}
t_{brems}=10^4\left[n/10^{11}\mbox{ cm}^{-3}\right]^{-1}\mbox{ s}
\end{equation}
is comparable to the IC loss time for the TeV electrons (\ref{eq:tickn}) and is 
shorter than the IC loss time for the X-ray emitting electrons (\ref{eq:tic}).
Thus, during the short period of escape from the dense equatorial disk (the
escape time is given by Eq. (\ref{eq:esc})) as much as 
\begin{equation}
L_\gamma/L_e\simeq t_{esc}/t_{brems}
\sim 10\%
\left[n_{\rm disk}/10^{11}\mbox{ cm}^{-3}\right]
\end{equation} 
of the power in relativistic
electrons, $L_e$, can be converted into the (bremsstrahlung) \gr\ luminosity,
$L_\gamma$. 

Below the electron energy $E_e\sim 350$~MeV the Coulomb energy loss, for
which the cooling time scale is given by 
\begin{equation}
t_{Coul}\simeq 3\times 10^3\left[n_{\rm disk}/10^{11}\mbox{ cm}^{-3}\right]^{-1}
\left[E_e/100\mbox{ MeV}\right]\mbox{ s}
\end{equation}
dominates over the bremsstrahlung loss. During the periods of the disk passage,
essentially 100\% of the power output in electrons with energies below
the "Coulomb break" 
\begin{equation}
\label{eq:coul}
E_{Coul}\simeq 30\left[n_{\rm disk}/10^{11}\mbox{ cm}^{-3}
\right]\mbox{ MeV}
\end{equation}
(estimated from the condition that the Coulomb loss time is
equal to the escape time, $t_{Coul}\sim t_{esc}$) will be channeled into the
heating of the disk, rather than on emission from the system. As a result, only
electrons with energies above $E_{Coul}$ can be injected into the compactified
PWN. 

\subsection{Protons.} 

GeV-TeV energy protons can loose their energy only in interactions with the 
protons from the stellar wind. The enhancement of the $pp$ interaction rate is 
expected during the pulsar passage through the dense equatorial disk of Be star.
The $pp$ interaction time
\begin{equation}
\label{eq:pp}
t_{pp}\simeq 1.6\times
10^4\left[n/10^{11}\mbox{ cm}^{-3}\right]^{-1}\mbox{ s}
\end{equation}
is comparable to the electron bremsstrahlung loss time (\ref{eq:brems}). Following the same way of
reasoning as in the case of bremsstrahlung, one can find that as much as 10\%
of the power $L_p$ contained in the
 PW protons can be channeled in the secondary particles
($\gamma$-rays, neutrinos, electrons, positrons) produced in $pp$ interactions.
The $\pi^0$ decay $\gamma$-rays carry away about 1/3 of the power output in $pp$
interactions. Thus, the "\gr\ efficiency" of $pp$ interactions is somewhat lower
than the efficiency of bremsstrahlung,
\begin{equation}
\label{eq:pp_eff}
L_{\gamma}/L_{p}\simeq 0.3 t_{esc}/t_{pp}\sim 3\%
\left[n_{\rm disk}/10^{11}\mbox{ cm}^{-3}\right]
\end{equation}
 However, if the PW is proton-dominated,
the luminosity of the \gr\ emission from $pp$ interactions can exceed the
bremsstrahlung luminosity.

\section{IC model of X-ray to TeV emission.} 

Taking into account that the seed photons for the IC scattering have energies of
about 10~eV (assuming the temperature of Be star $T\simeq 2\times 10^4$~K), one
can find that the IC emission  from electrons of the energy $E_e$ peaks at
\begin{equation}
\label{eq:eic}
\epsilon_{IC}\simeq 4
\left[E_e/10\mbox{ MeV}\right]^2\mbox{ keV}
\end{equation}
The energy of the upscattered photons becomes approximately equal to the energy
of electrons at 
\begin{equation}
\epsilon_{IC, \rm (T\rightarrow KN)}\simeq 30\mbox{ GeV}
\end{equation} 
(the transition to the
Klein-Nishina regime). If the spectrum of electrons is a simple powerlaw with
the spectral index $p_e$ ($dN_e/dE\sim E^{-p_e}$), the IC spectrum below and
above the Thompson -- Klein-Nishina break is, respectively, 
$dN_\gamma/dE\sim E^{(p_e+1)/2}$ and $dN_\gamma/dE\sim E^{-p_e+1}\ln E$. 
The IC emission in the 10-100~GeV energy band is characterized by one more
spectral feature. Namely, the IC cooling time of the 10-100~GeV electrons is
comparable to the escape time from the compact region with a dense photon
background. Estimating the energy of the cooling break in the IC emission 
spectrum from the condition 
$t_{esc}\simeq t_{IC,\rm\ (T)}$ one finds
$\epsilon_{IC,\rm\ cool}\simeq 4\left[R/10^{13}\mbox{cm}\right]
\mbox{ GeV}$. Taking into account the coincidence of the cooling break energy 
with the energy of transition to the Klein-Nishina regime  one can not expect
to detect the conventional steepening of the IC spectrum above the cooling break
because of the reduced efficiency of the IC scattering in the Klein-Nishina
regime. One more complication of the detailed calculation of the IC emission
spectrum in the GeV-TeV energy band is that in order to explain the observed behaviour of the TeV lightcurve
during the periastron passage within the IC model one has to assume that either 
additional non-radiative cooling mechanism dominates electron energy loss close
to the periastron, or a cut-off in the electron spectrum at sub-TeV energies is
present \cite{khangulyan06}. The combined effect of the above mentioned difficulties makes the
detailed predictions for the IC spectrum in the GeV-TeV band quite uncertain and
we do not attempt the detailed fit of the observed spectrum in this band. 
Instead we concentrate of the attempt to fit the general shape of the
spectral energy distribution in the X-ray to TeV \gr band within the IC model.

\begin{figure}
\includegraphics[width=0.8\columnwidth,angle=0]{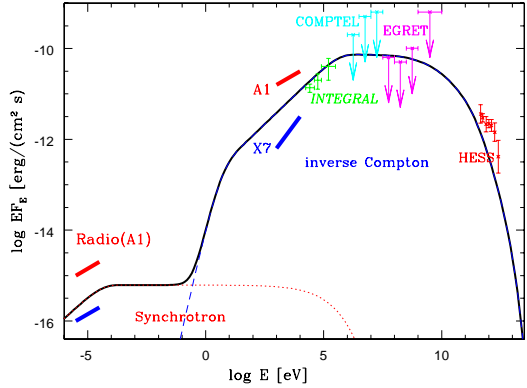}
\caption{IC model fit for the X-ray to TeV spectrum of the source. Radio
emission is synchrotron from electrons which produce X-ray IC flux.
}
\label{fig:SEC}
\end{figure}

Fig. \ref{fig:SEC} shows an example of the fit the the spectrum of \psrb\ in IC 
model for X-ray to TeV emission. One can see that EGRET upper limit on the flux
from the system requires the presence of a break in the IC spectrum at the
energies $E\sim 1$~MeV. In the model fit shown in the
Figure, the electron spectrum below the break at $E_e=100$~MeV has the spectral
index $p_e=2$, while above the break the spectrum steepens to $p_e+1=3$.
It is clear that the overall shape of the IC 
spectrum in the keV to TeV energy band agrees well with the data. 

The energy of the break in the electron spectrum ($\sim 100$~MeV)
is close to the energy of the Coulomb break given by Eq. (\ref{eq:coul}).  
As it was discussed above,  electrons with energies
below $E_{Coul}$ loose all their energy via the severe Coulomb loss before
they are able to escape from the dense equatorial disk of Be star to the less
dense PWN. 
As a result, regardless of the initial injection spectrum of electrons
from the PW, the spectrum of electrons injected in the compact PWN 
has a low-energy cut-off at the energy $\sim E_{Coul}$. The IC cooling of 
electrons in the PWN leads to the formation of the characteristic powerlaw 
tail of electron
distribution below $E_{Coul}$ with $p_e=2$. The electron spectrum above the
energy $E_{Coul}$ (assumed to be a powerlaw with the spectral index $p_e=3$ in
the model fit of Fig. \ref{fig:SEC}) is determined by the balance of acceleration and energy losses
in the pulsar/stellar wind shock region.

Electrons responsible for the X-ray IC emission   produce synchrotron radiation
in radio band at the characteristic frequency 
\begin{equation}
\epsilon_S\simeq 1.5\left[B/0.1\mbox{ G}\right]\left[E_e/30\mbox{ MeV}\right]^2
\mbox{ GHz}
\end{equation}
 The ratio of the synchrotron to IC
luminosity is given by the ratio of the energy densities of the magnetic field
and radiation,
\begin{equation}
L_S/L_{IC}=2\times 10^{-4}\left[B/0.1\mbox{
G}\right]^2\left[R_{\rm PWN}/10^{13}\mbox{ cm}\right]^2
\end{equation}
The radio luminosity of the system is some 4 orders of magnitude lower than the
X-ray luminosity.  This imposes a restriction on the possible strength of
magnetic field in the X-ray emission region,
\begin{equation}
\label{eq:bmax}
B\le 0.1\left[R_{\rm PWN}/10^{13}\mbox{ cm}\right]^{-1}\mbox{ G}
\end{equation}
In the model fit of Fig. \ref{fig:SEC} we have chosen the magnetic field
strength $B=0.03$~G and assumed the size of X-ray / radio emission region 
$R_{\rm PWN}\sim 3\times 10^{13}$~cm.

\section{Alternative mechanisms  of TeV \gr\ emission.} 
\label{sec:last}
\subsection{Bremsstrahlung.} 

The IC model for the keV-to-TeV spectrum has a difficulty to explain
the observed correlation of the radio, X-ray and TeV emission because the 
cooling and escape times of electrons emitting IC radiation in
X-ray and TeV bands are different. Since the non-pulsed radio emission from the
system is most probably related to the passage of the pulsar through the disk of
Be star, an explanation of the observed correlation requires a physical
mechanism which would explain the increase of the TeV flux during the disk
passage. 
At least two mechanisms of interaction of the pulsar wind with the Be star disk
can lead to the increase of TeV emission: bremsstrahlung and proton-proton
interactions. 

\begin{figure}
\includegraphics[width=0.8\columnwidth,angle=0]{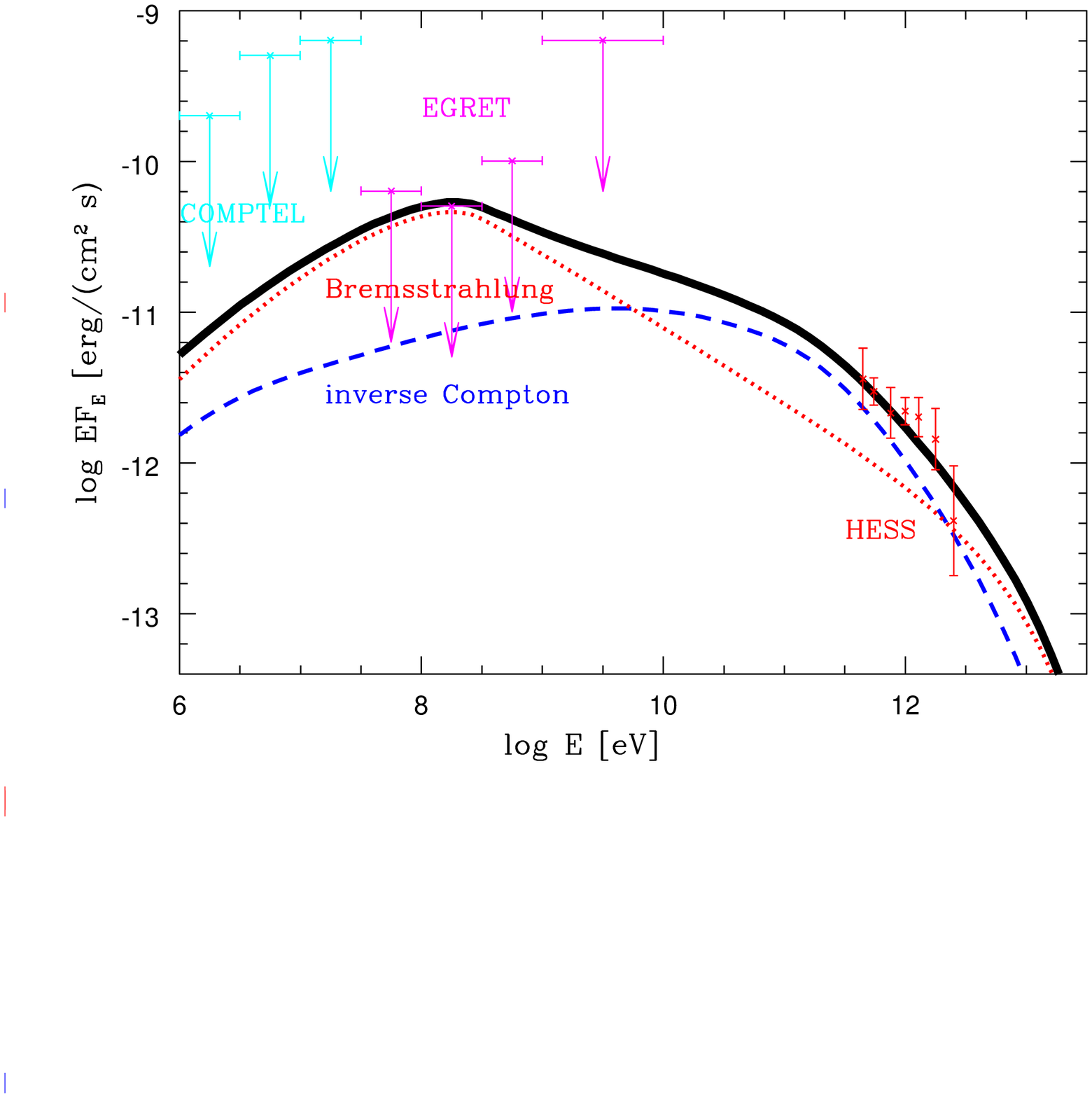}
\caption{Comparison of bremsstrahlung and IC contributions to
the \gr\ flux. See text for the values of parameters used for the model fits.
}
\label{fig:BREMS}
\end{figure}

As it is discussed above, the bremsstrahlung cooling time in the
dense Be star disk (\ref{eq:brems}) can be comparable to the IC cooling time
both for the highest energy electrons above TeV and for electrons with energies
below 1~GeV (see Eqs. (\ref{eq:tic}), (\ref{eq:tickn})). The bremsstrahlung
cooling time can be comparable to the escape time from the compact equatorial
disk so that up to 10\% of the power of the pulsar wind can be emitted in the
form of bremsstrahlung radiation. 
Fig. \ref{fig:BREMS} shows the fit for the $\gamma$-ray spectrum of the system 
with a combination of IC and bremsstrahlung emission. The electron
spectrum is supposed to be a cut-off powerlaw with the spectral index $p_e
=2.5$ and cut-off energy $E_{cut}=20$~TeV. Note that the EGRET upper limit 
imposes a restriction on the spectrum of electrons because the bremsstrahlung
spectrum has the photon index $\Gamma_{ph}\simeq p_e$. Assuming that
the electron spectrum continues to lower energies without a break would violate
the EGRET bound on the flux. A break at the energy $E\simeq 350$~MeV was assumed
in the electron spectrum in the model fit of Fig. \ref{fig:BREMS}. 
The break at this particular energy is naturally expected in the bremsstrahlung
scenario, because below this energy the ionization loss dominates over the
bremsstrahlung loss, which leads to the hardening of the electron spectrum 
by $\Delta p_{e}\simeq 1$ at
low energies.

\subsection{$pp$ interactions.} 

\begin{figure}
\includegraphics[width=0.8\columnwidth,angle=0]{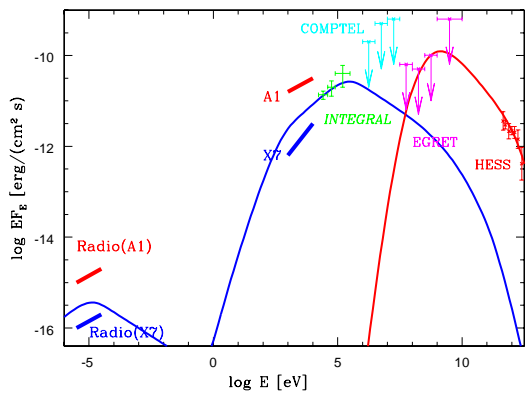}
\caption{Broad band spectrum in the $pp$ model. \gr\ emission is produced via
neutral pion decays (red line) while radio and X-ray emission are, respectively,
synchrotron and IC radiation from electrons/positrons produced in the charged
pion decays.
}
\label{fig:PP}
\end{figure}

If the pulsar wind is proton-loaded, interactions of the pulsar wind
protons with the protons from the dense Be star disk provide an additional
source the TeV \gr\ emission.
As is is discussed above, the "\gr\ efficiency" of $pp$
interactions is a factor of several lower than that of bremsstrahlung, 
but the relative contributions of bremsstrahlung and $pp$ interactions into the
\gr\ emission depend on the proton-to-electron ratio of the PW. 
 If the pulsar wind is mostly
proton loaded, the $\pi^0$ decay emission can dominate over the $\gamma$-ray
emission from the pulsar wind electrons. An example of the fit to the TeV \gr
spectrum within the $pp$ model is shown in Fig. \ref{fig:PP}. We have assumed
a powerlaw spectrum of protons with the spectral index $p_p=2.6$ for the model
fit. Similarly to the case of bremsstrahlung, the EGRET upper limit on the flux
imposes a restriction on the spectrum of the protons at several GeV energies. 
However, contrary to the bremsstrahlung case, the spectrum of $\pi^0$ decays \gr
emission has a low energy cut off below GeV energy and the restriction on the
spectrum of the protons is rather weak. In fact, in the model fit of Fig.
\ref{fig:PP} the proton spectrum is described by a single powerlaw from GeV to
TeV energies. It is important to note that in this case the luminosity of the
source in the 10~GeV energy band is higher than in the TeV
band.

If bremsstrahlung or pion decay emission dominate the TeV flux, 
the TeV luminosity is just proportional to the density of the Be star disk at the location of the pulsar. This means that in this case the parameters of the Gaussian that approximately fits the TeV lightcurve (Fig. \ref{gauss}) should be identified with the equatorial plane position ($\theta_0$) and width ($\Delta\theta_0$) of the disk. Within such interpretation the disk appears to be ``thick'' 
and the calculation of the phases of disappearance/reappearance of the pulsed emission has to be done 
by calculating the column density of the disk along the line of sight for different positions of the pulsar. Fixing the phases of disappearance/reappearance of the pulsed emission to the known values, one can obtain a constraint on the radial density profile of the disk and/or inclination of the disk w.r.t. the line of sight. However, if inverse Compton emission gives a contribution comparable to 
the pion decay or bremsstrahlung, the derivation of the disk parameters from the shape of the TeV lightcurve is not possible.

The most important feature of the model with $pp$ interactions is that it can
explain the broad band spectrum of the system from radio to TeV energy band.
The idea is that
the synchrotron and IC emission from
the secondary electrons produced in the decays on charged pions is emitted in
the radio and X-ray bands, respectively, while the bremsstrahlung emission 
from the secondary electrons, emitted in the $\gamma$-ray band gives a
sub-dominant contribution, compared to the $\pi^0$ decay emission. 
Fig. \ref{fig:PP} shows an example of the fit to the broad band spectrum of the
system within the $pp$ interactions model. Since the emission in radio, X-ray
and TeV bands is produced via one and the same process ($pp$ interactions), the
observed correlation of the radio, X-ray and TeV flux is naturally explained. 
Besides, the  observed hardening 
of the X-ray spectrum during a several-day period following the moment of the
entrance of the pulsar to the disk of Be star is explained by the low energy
cut-off at $\sim 100$~MeV in the spectrum of secondary electrons. Such a cut off
arises (a) because of the kinematics of the pion decays and (b) 
because of the efficient Coulomb cooling of electrons with energies below
100~MeV
during the escape from the Be star disk.

The $pp$ interaction scenario is attractive because of one more reason:  
in this model the  overall energy balance of the system is evident. Indeed, 
in the "purely electronic" models it is not clear why the system is "radiatively
inefficient": the spin-down luminosity of the pulsar is $\simeq 10^{36}$~erg/s,
but the bolometric luminosity is just $L< 10^{34}$~erg/s, which accounts for 
no more than
one percent of the spin-down luminosity. To the contrary, within "protonic"
model one has to assume that proton-loaded PW carries a significant fraction of
the spin-down power.
As it is explained above, in the $pp$
model the efficiency of conversion of the power contained in the protons
into the $\gamma$-ray emission is  several percents, (see Eq. (\ref{eq:pp_eff}))
 which explains the 
\gr luminosity $L_\gamma\sim 10^{34}$ erg/s. 

The $pp$ model can be readily tested with the future observations of the system
in the 10~GeV energy band with {\it GLAST}. Indeed, from Fig. \ref{fig:PP} one
can see that in the $pp$ model the EGRET upper limit on the flux at 10~GeV
should be close to the actual level of the $\gamma$-ray flux from the system.
This means that the detection of the system during the periastron passage 
with a more sensitive instrument, like GLAST should not be a problem.

\begin{acknowledgements}
We would like to thank F.Aharonian for the fruitful discussions of the subject of the paper.
\end{acknowledgements}

\end{document}